\documentclass[a4paper]{rnti}
\pdfoutput=1
\usepackage{amsmath,amssymb}
\usepackage{theorem}
\usepackage[dvips]{graphicx}
\usepackage{algorithm}
\usepackage{algorithmic}
\usepackage{lscape}
\usepackage{ifpdf}
\usepackage{color}
\ifpdf
\else

\fi
\def \etal {{\it et al.~}} 
\newcommand{\bftt}[1]{{\textbf{\texttt{#1}}}} 
\newcommand{\set}[1]{\left\{#1\right\}} 
\floatname{algorithm}{Algorithme}

\titrecourt{S\'{e}lection simultan\'{e}e d'index et de vues mat\'{e}rialis\'{e}es} \nomcourt{N.
Maiz et al.}

\titre{S\'{e}lection simultan\'{e}e d'index et de vues mat\'{e}rialis\'{e}es}

\auteur{Nora Maiz, Kamel Aouiche et J\'{e}r\^{o}me Darmont}

\affiliation{
    Laboratoire ERIC, Universit\'{e} Lumi\`{e}re Lyon 2\\ 5 avenue Pierre Mend\`{e}s-France \\69676 Bron
    Cedex \\
          \{nora.maiz, kamel.aouiche, jerome.darmont\}@eric.univ-lyon2.fr\\
          \http{http://eric.univ-lyon2.fr}\\}

\resume{ Les index et les vues mat\'{e}rialis\'{e}es sont des structures physiques qui
acc\'{e}l\`{e}rent l'acc\`{e}s aux donn\'{e}es d'un entrep\^{o}t. Ces structures engendrent
cependant une surcharge de maintenance. Par ailleurs, elles partagent le m\^{e}me
espace disque. Les travaux existants dans le domaine de la s\'{e}lection d'index et
de vues mat\'{e}rialis\'{e}es traitent ces deux structures de mani\`{e}re isol\'{e}e. Dans cet
article, nous couplons au contraire la s\'{e}lection d'index et de vues
mat\'{e}rialis\'{e}es de fa\c{c}on \`{a} prendre en compte les interactions entre ces
structures de donn\'{e}es et \`{a} permettre un partage efficace de l'espace de
stockage commun qui leur est allou\'{e}. Pour cela, nous avons d\'{e}velopp\'{e} des
mod\`{e}les de co\^{u}t qui \'{e}valuent le b\'{e}n\'{e}fice de la mat\'{e}rialisation de vue et de
l'indexation. Ces mod\`{e}les de co\^{u}t nous permettent, gr\^{a}ce \`{a} un algorithme
glouton, de s\'{e}lectionner une configuration pertinente d'index et de vues
mat\'{e}rialis\'{e}es. Nos exp\'{e}rimentations montrent que notre strat\'{e}gie se r\'{e}v\`{e}le
meilleure que celles qui op\`{e}rent une s\'{e}lection isol\'{e}e des index et des vues
mat\'{e}rialis\'{e}es.}

\summary{ Indices and materialized views are physical structures that
accelerate data access in data warehouses. However, these data structures
generate some maintenance overhead. They also share the same storage space. The
existing studies about index and materialized view selection consider these
structures separately. In this paper, we adopt the opposite stance and couple
index and materialized view selection to take into account the interactions
between them and achieve an efficient storage space sharing. We develop cost
models that evaluate the respective benefit of indexing and view
materialization. These cost models are then exploited by a greedy algorithm to
select a relevant configuration of indices and materialized views. Experimental
results show that our strategy performs better than the independent selection
of indices and materialized views. }

\begin{document}

\section{Introduction}

Les entrep\^{o}ts de donn\'{e}es sont g\'{e}n\'{e}ralement mod\'{e}lis\'{e}s selon un sch\'{e}ma en \'{e}toile
contenant une table de faits centrale volumineuse et un certain nombre de
tables dimensions repr\'{e}sentant les descripteurs des faits~\cite{inm02bui,kim02dat}. La table de faits contient des cl\'{e}s \'{e}trang\`{e}res vers les cl\'{e}s
primaires des tables dimensions, ainsi que des mesures num\'{e}riques. Avec ce type
de mod\`{e}le, une requ\^{e}te d\'{e}cisionnelle n\'{e}cessite une ou plusieurs jointures entre
la table de faits et les tables dimensions. De plus, le sch\'{e}ma de l'entrep\^{o}t
peut comporter des hi\'{e}rarchies au niveau des dimensions (sch\'{e}ma en flocon de
neige), ce qui entra\^{\i}ne des jointures additionnelles. Ces jointures sont tr\`{e}s
co\^{u}teuses en terme de temps de calcul. Ce co\^{u}t devient prohibitif lorsque les
jointures op\`{e}rent sur de tr\`{e}s grands volumes de donn\'{e}es. Il est alors crucial
de le r\'{e}duire.

Les vues mat\'{e}rialis\'{e}es et les index sont des structures
physiques qui permettent de r\'{e}duire le temps d'ex\'{e}cution des
requ\^{e}tes en pr\'{e}calculant les jointures et en offrant un
acc\`{e}s direct aux donn\'{e}es.
Cependant, lors du
rafra\^{\i}chissement de l'entrep\^{o}t de donn\'{e}es, ces structures doivent \'{e}galement \^{e}tre mises
\`{a} jour, ce qui engendre une surcharge pour le syst\`{e}me. Par ailleurs, index et vues mat\'{e}rialis\'{e}es
partagent le m\^{e}me espace  de stockage. Il est donc judicieux de ne cr\'{e}er que les
plus pertinents.

Les travaux existants dans le domaine de la s\'{e}lection d'index
et/ou de vues mat\'{e}ria\-lis\'{e}es traitent ces deux structures
de mani\`{e}re isol\'{e}e ou s\'{e}quentielle. Dans cet article,
nous proposons une nouvelle strat\'{e}gie qui op\`{e}re une
s\'{e}lection simultan\'{e}e des vues mat\'{e}rialis\'{e}es et des
index afin de prendre en compte les interactions entre eux.

Dans un premier temps, nous exploitons des strat\'{e}gies de
s\'{e}lection isol\'{e}e des vues mat\'{e}rialis\'{e}es et des
index, qui nous fournissent un ensemble d'index et de vues
candidats. Nous calculons ensuite gr\^{a}ce \`{a} des mod\`{e}les de
co\^{u}t le b\'{e}n\'{e}fice potentiel de chaque index et de chaque
vue mat\'{e}rialis\'{e}e, en prenant en compte les interactions
possibles entre ces deux types de structures. Finalement, un
algorithme glouton nous permet de s\'{e}lectionner simultan\'{e}ment
les vues mat\'{e}rialis\'{e}es et les index les plus pertinents.

Cet article est organis\'{e} comme suit. La Section~\ref{sec:etat_art} est
consacr\'{e}e \`{a} l'\'{e}tat de l'art de ce domaine de recherche. La
Section~\ref{sec:strategy_sim}  pr\'{e}sente globalement notre strat\'{e}gie de
s\'{e}lection simultan\'{e}e de vues mat\'{e}rialis\'{e}es et d'index. Nous d\'{e}taillons ensuite
nos mod\`{e}les de co\^{u}t dans la Section~\ref{sec:cost_modele_sim}, puis le calcul
des b\'{e}n\'{e}fices d'indexation et de mat\'{e}rialisation de vues dans la
Section~\ref{sec:calcul_benefice} . Nous pr\'{e}sentons dans la
Section~\ref{sec:algo_sim} notre algorithme glouton de s\'{e}lection simultan\'{e}e de
vues mat\'{e}rialis\'{e}es et d'index. Les premi\`{e}res exp\'{e}rimentations que nous avons
men\'{e}es pour valider notre approche sont pr\'{e}sent\'{e}es dans la
Section~\ref{sec:exp_sim}. Nous concluons finalement cet article et \'{e}voquons
nos perspectives de recherche dans la Section~\ref{sec:conclusion_sim}.

\section{\'{E}tat de l'art}
\label{sec:etat_art}

Le probl\`{e}me de s\'{e}lection d'index et/ou de vues mat\'{e}rialis\'{e}es consiste \`{a}
construire une configuration d'index et/ou de vues mat\'{e}rialis\'{e}es optimisant le
co\^{u}t d'ex\'{e}cution d'une charge donn\'{e}e, suppos\'{e}e repr\'{e}sentative. Cette optimisation peut \^{e}tre r\'{e}alis\'{e}e
sous certaines contraintes, comme l'espace de stockage allou\'{e} aux index et aux
vues \`{a} s\'{e}lectionner.

Plus formellement, si $O=\{o_{1},...,o_{n}\}$ est un ensemble
d'objets (index candidats, vues mat\'{e}rialis\'{e}es candidates ou
index sur les vues), $Q~=~\{q_{1},...,q_{m}\}$ l'ensemble des
requ\^{e}tes de la charge et $S$ la taille de l'espace disque
allou\'{e} par l'administrateur pour stocker les objets \`{a}
s\'{e}lectionner, alors il faut trouver une configuration d'index et
de vues mat\'{e}rialis\'{e}es $Config$ tel que :
\begin{itemize}

\item le co\^{u}t d'ex\'{e}cution $C$ des requ\^{e}tes de la charge soit minimal,
c'est-\`{a}-dire~:  $$C_{/Config}(Q) = Min \left(C_{/O}(Q)\right);$$

\item l'espace de stockage des index et des vues de $Config$ ne d\'{e}passe pas $S$, c'est-\`{a}-dire~:
 $$\sum_{o_{i} \in Config} taille(o_{i}) \leq S.$$
\end{itemize}

Les probl\`{e}mes de s\'{e}lection  d'index et de vues mat\'{e}rialis\'{e}es  sont connus pour
\^{e}tre NP-complets \cite{com78dif,gup99sel-a}. De ce fait, il n'existe pas
d'algorithme qui propose une solution optimale en un temps fini. Plusieurs
travaux de recherche proposent des solutions proches de la solution optimale en
utilisant des heuristiques r\'{e}duisant la complexit\'{e} du probl\`{e}me.

Les travaux traitant la s\'{e}lection d'index
\cite{fra92ada,cho93ind,cho93sel,val00db2,gol02ind,dageville04} et  la
s\'{e}lection de vues \cite{gup99sel-a,gup99sel-b,bar03sel,smi04wav,gup05sel}
s'orientent dans leur majorit\'{e} vers une s\'{e}lection s\'{e}quentielle des vues et des
index sur les vues, ou vers une s\'{e}lection isol\'{e}e des index ou des vues
mat\'{e}rialis\'{e}es. Cependant, les index et les vues mat\'{e}rialis\'{e}es sont
fondamentalement des structures physiques similaires~\cite{agr00aut}. En effet,
les deux structures sont redondantes, acc\'{e}l\`{e}rent le temps d'ex\'{e}cution des
requ\^{e}tes, partagent la m\^{e}me ressource de stockage et impliquent une surcharge
de maintenance pour le syst\`{e}me suite aux mises \`{a} jour des donn\'{e}es. Les vues et
les index peuvent alors \^{e}tre en interaction. La pr\'{e}sence d'un index sur une vue
mat\'{e}rialis\'{e}e peut en effet rendre celle-ci plus ``attractive'' et \textit{vice
versa}.

Peu de travaux se sont port\'{e}s sur la s\'{e}lection simultan\'{e}e des vues mat\'{e}rialis\'{e}es
et des index. Agrawal~\etal ont propos\'{e} trois alternatives pour l'\'{e}num\'{e}ration
conjointe de l'espace des index et des vues mat\'{e}rialis\'{e}es \cite{agr00aut}. La
premi\`{e}re alternative, d\'{e}not\'{e}e MVFIRST, tend \`{a} s\'{e}lectionner les vues
mat\'{e}rialis\'{e}es en premier, puis les index pour une charge donn\'{e}e en pr\'{e}sence des
vues pr\'{e}alablement s\'{e}lectionn\'{e}es. La deuxi\`{e}me alternative, d\'{e}not\'{e}e INDFIRST,
s\'{e}lectionne en premier les index, puis les vues. La troisi\`{e}me alternative,
d\'{e}not\'{e}e \textit{joint enumeration}, traite la s\'{e}lection des index, des vues
mat\'{e}rialis\'{e}es et des index sur ces vues en une seule it\'{e}ration. Les auteurs
affirment qu'elle est plus efficace que les deux premi\`{e}res.

Bellatreche~\etal ont trait\'{e} le probl\`{e}me de distribution de l'espace de
stockage entre les vues mat\'{e}rialis\'{e}es et les index de mani\`{e}re it\'{e}rative afin de
minimiser le co\^{u}t total d'ex\'{e}cution des requ\^{e}tes d'une charge
donn\'{e}e~\cite{bel00eff}.  Un ensemble de vues et d'index est d\'{e}sign\'{e} comme une
solution initiale au probl\`{e}me de s\'{e}lection d'index et de vues. L'approche
reconsid\`{e}re it\'{e}rativement la solution initiale dans le but de r\'{e}duire davantage
le co\^{u}t d'ex\'{e}cution des requ\^{e}tes en redistribuant l'espace de stockage entre
les vues et les index. Elle s'appuie sur une comp\'{e}tition perp\'{e}tuelle entre deux
agents, l'espion des index et l'espion des vues. L'espion des index
(respectivement, des vues) vole de l'espace r\'{e}serv\'{e} pour stocker les vues
(respectivement, les index). L'espace ainsi r\'{e}cup\'{e}r\'{e} est utilis\'{e} pour cr\'{e}er
d'autres index \`{a} la place des vues \'{e}lagu\'{e}es, suivant des politiques de
remplacement. L'op\'{e}ration est valid\'{e}e si le co\^{u}t d'ex\'{e}cution des requ\^{e}tes est
r\'{e}duit. La s\'{e}lection d'index et de vues commence par appliquer l'espion qui
r\'{e}duit le plus le co\^{u}t des requ\^{e}tes. Le processus de s\'{e}lection s'arr\^{e}te
lorsqu'il n'y a plus de r\'{e}duction du co\^{u}t des requ\^{e}tes.

Finalement, Rizzi et Saltarelli ont propos\'{e} une approche qui
d\'{e}termine \textit{a priori} un compromis entre l'espace de
stockage allou\'{e} aux index et aux vues mat\'{e}rialis\'{e}es en
se basant sur les requ\^{e}tes de la charge~\cite{vie03riz}.
L'id\'{e}e de Rizzi et Saltarelli est que le facteur cl\'{e} dans
l'optimisation des performances des requ\^{e}tes est leur niveau
d'agr\'{e}gation, d\'{e}fini par la liste des attributs de la clause
\bftt{Group by}, et la s\'{e}lectivit\'{e} des attributs
pr\'{e}sents dans les clauses \bftt{Having} et \bftt{Where}. En
effet, la mat\'{e}rialisation offre un grand b\'{e}n\'{e}fice aux
requ\^{e}tes comportant des agr\'{e}gations de granularit\'{e}
grossi\`{e}re (nombre faible d'attributs dans la clause \bftt{Group
by}) car elles g\'{e}n\`{e}rent peu de groupes dans un grand nombre
de n-uplets et, par cons\'{e}quent, l'acc\`{e}s \`{a} une petite vue
est moins co\^{u}teux que l'acc\`{e}s aux tables de base. D'autre
part, les index donnent leur meilleur b\'{e}n\'{e}fice avec des
requ\^{e}tes contenant des attributs dont la s\'{e}lectivit\'{e} est
\'{e}lev\'{e}e car elles s\'{e}lectionnent peu de n-uplets et, par
cons\'{e}quent, l'acc\`{e}s \`{a} un nombre \'{e}lev\'{e} de
n-uplets inutiles est \'{e}vit\'{e}. Les requ\^{e}tes avec des
agr\'{e}gations fines et de fortes s\'{e}lectivit\'{e}s encouragent
l'indexation. En revanche, les requ\^{e}tes avec des
agr\'{e}gations grossi\`{e}res et de faibles s\'{e}lectivit\'{e}s
encouragent la mat\'{e}rialisation.

Le d\'{e}faut que nous identifions dans les travaux de Bellatreche~\etal et de
Rizzi et Saltarelli est qu'ils ne prennent pas en compte les interactions entre
les vues mat\'{e}rialis\'{e}es et les index. En effet, les deux types de structures y
sont s\'{e}lectionn\'{e}s de mani\`{e}re concurrente et non conjointe. Or, des index sur
les vues mat\'{e}rialis\'{e}es d\'{e}j\`{a} s\'{e}lectionn\'{e}es peuvent s'av\'{e}rer des options tr\`{e}s
int\'{e}ressantes. Notre approche s'apparente donc \`{a} celle d'Agrawal~\etal
(\textit{joint enumeration}). Cependant, ses auteurs ne donnent malheureusement
aucun d\'{e}tail sur son fonctionnement, ce qui rend toute comparaison (y compris
exp\'{e}rimentale) impossible.

\section{S\'{e}lection simultan\'{e}e d'index et de vues \\mat\'{e}rialis\'{e}es}
\label{sec:strategy_sim}

\begin{figure}[hbt]
{\centering \resizebox*{0.8\textwidth}{!}{\includegraphics{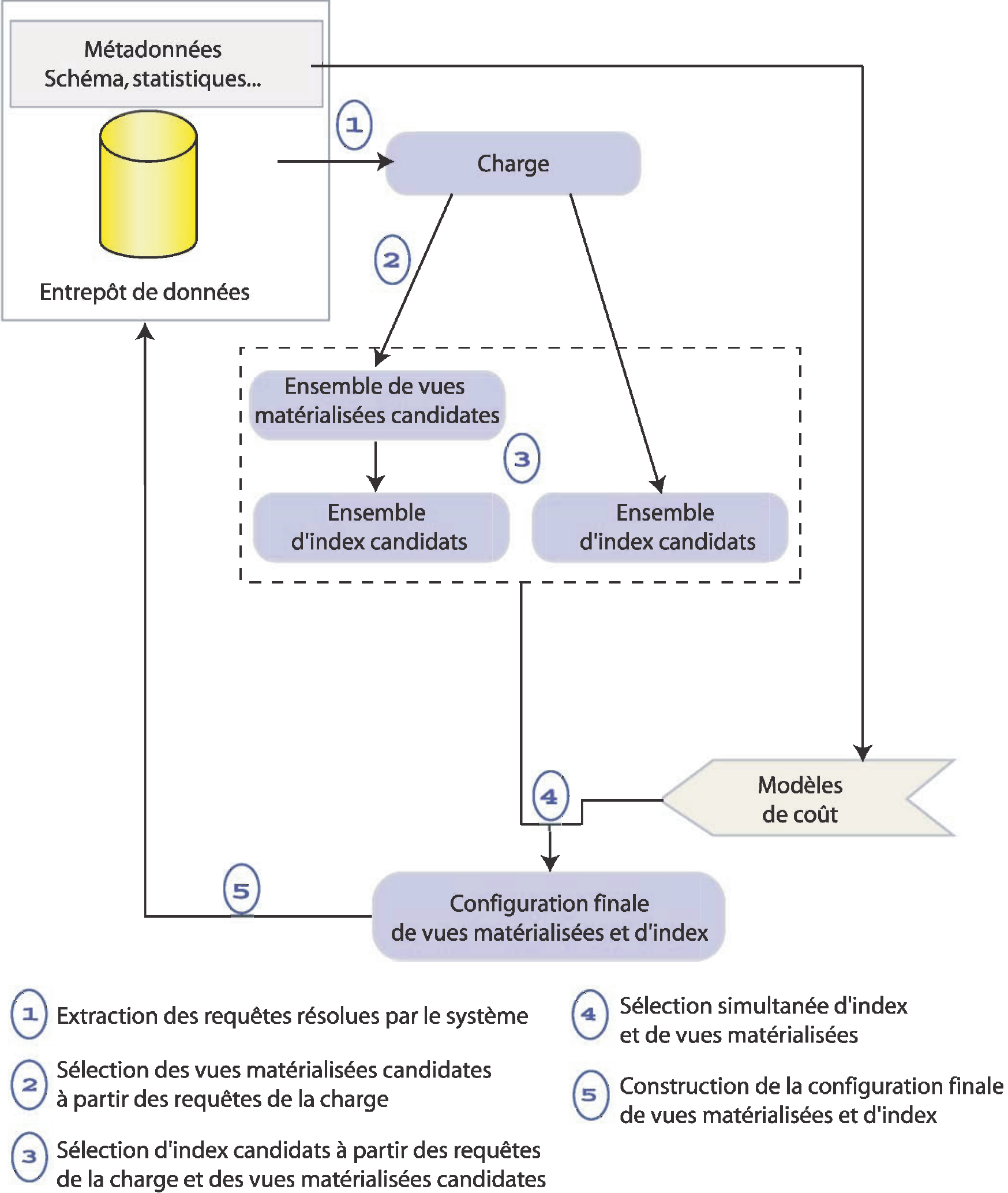}}
\par}
\caption{Principe de notre s\'{e}lection simultan\'{e}e d'index et de
vues mat\'{e}rialis\'{e}es} \label{fig:archi_index_vue}
\end{figure}

Le principe g\'{e}n\'{e}ral de notre strat\'{e}gie de s\'{e}lection
simultan\'{e}e d'index et de vues mat\'{e}rialis\'{e}es est
repr\'{e}sent\'{e} \`{a} la Figure~\ref{fig:archi_index_vue}.
Rappelons qu'une vue mat\'{e}rialis\'{e}e est une requ\^{e}te
nomm\'{e}e dont les donn\'{e}es sont stock\'{e}es sur disque sous la
forme d'une table. La s\'{e}lection d'index peut donc se faire sur
les tables de base ainsi que sur les vues mat\'{e}rialis\'{e}es.
Nous proc\'{e}dons comme suit pour proposer une configuration
d'index et de vues mat\'{e}rialis\'{e}es pertinents :

\begin{itemize}

\item extraction d'une charge des requ\^{e}tes repr\'{e}sentative,
\item construction de l'ensemble des vues mat\'{e}rialis\'{e}es candidates \`{a} partir de la charge,
\item construction de l'ensemble des index candidats \`{a} partir de la charge et des vues mat\'{e}rialis\'{e}es candidates,
\item s\'{e}lection simultan\'{e}e d'index et de vues mat\'{e}rialis\'{e}es,
\item construction de la configuration finale d'index et de vues mat\'{e}rialis\'{e}es.

\end{itemize}

L'extraction  de la charge en entr\'{e}e de notre approche s'effectue \`{a} partir du
journal des requ\^{e}tes ex\'{e}cut\'{e}es sur les donn\'{e}es de l'entrep\^{o}t. La charge que
nous consid\'{e}rons est un ensemble de requ\^{e}tes de projection, s\'{e}lection et
jointure. De telles requ\^{e}tes sont compos\'{e}es d'op\'{e}rations de jointures, de
pr\'{e}dicats de s\'{e}lection et d'op\'{e}rations d'agr\'{e}gation. Nous appliquons ensuite
une strat\'{e}gie de s\'{e}lection isol\'{e}e de vues mat\'{e}rialis\'{e}es que nous avons
d\'{e}velopp\'{e}e et qui est bas\'{e}e sur la classification non supervis\'{e}e des requ\^{e}tes
\cite{aou05tec}. Cela permet de construire un ensemble de vues candidates
pertinent pour la charge. La classification des requ\^{e}tes peut \^{e}tre vue comme
une sorte de compression de la charge~\cite{chaudhuri02}. Cela permet d'assurer
la scalabilit\'{e} de notre approche. En effet, les charges ont tendance \`{a} \^{e}tre
volumineuses et leur co\^{u}t de traitement est par cons\'{e}quence important. Au lieu
de r\'{e}aliser l'optimisation directement \`{a} partir de la charge, il est plus
judicieux de le faire \`{a} partir d'une charge compress\'{e}e qui conserve les
relations existant entre les requ\^{e}tes de la charge initiale.

Nous avons \'{e}galement d\'{e}velopp\'{e} une strat\'{e}gie de s\'{e}lection isol\'{e}e d'index bas\'{e}e
sur la recherche de motifs fr\'{e}quents ferm\'{e}s~\cite{aou03fre,aou05aut},
qui permet de construire un ensemble d'index pertinent pour les requ\^{e}tes de la
charge et les vues mat\'{e}rialis\'{e}es candidates g\'{e}n\'{e}r\'{e}es \`{a} l'\'{e}tape pr\'{e}c\'{e}dente.
Notons que, notre approche \'{e}tant modulaire, nous pourrions utiliser toute autre
m\'{e}thode de s\'{e}lection isol\'{e}e d'index ou de vues mat\'{e}rialis\'{e}es. Finalement, nous
appliquons notre strat\'{e}gie de s\'{e}lection simultan\'{e}e de vues mat\'{e}rialis\'{e}es et
d'index, que nous d\'{e}taillons dans les sections suivantes.

Afin d'illustrer notre propos, nous nous basons sur la charge de requ\^{e}tes de la
Figure~\ref{fig:charge_sim} et les vues mat\'{e}rialis\'{e}es et les index candidats
qui en d\'{e}coulent (Figures~\ref{fig:vue_sim}~et~\ref{fig:index_sim},
respectivement). Nous mod\'{e}lisons les relations existant entre les requ\^{e}tes, les
vues mat\'{e}rialis\'{e}es
 et les index candidats \`{a} l'aide de trois matrices : \emph{requ\^{e}tes-vues},
\emph{requ\^{e}tes-index} et \emph{vues-index}, que nous d\'{e}crivons dans les sections suivantes.

\subsection{Matrice requ\^{e}tes-vues}

La matrice requ\^{e}tes-vues mod\'{e}lise les relations
entre les requ\^{e}tes de la charge et les vues
mat\'{e}rialis\'{e}es qui en sont extraites, c'est-\`{a}-dire
les vues exploit\'{e}es par au moins une requ\^{e}te de la charge. Cette
matrice peut \^{e}tre vue comme le r\'{e}sultat de la
r\'{e}\'{e}criture des requ\^{e}tes de la charge en fonction des
vues mat\'{e}rialis\'{e}es. Les lignes et les colonnes de cette
matrice sont les requ\^{e}tes de la charge et les vues
mat\'{e}rialis\'{e}es recommand\'{e}es par notre strat\'{e}gie de
s\'{e}lection de vues, respectivement. Le terme g\'{e}n\'{e}ral de
la matrice est \'{e}gal \`{a} un si une requ\^{e}te donn\'{e}e
exploite une vue et \`{a} z\'{e}ro sinon. Le Tableau~\ref{tab:req_vue}
illustre un exemple de matrice requ\^{e}tes-vues
compos\'{e}e de huit requ\^{e}tes et de neuf vues mat\'{e}rialis\'{e}es recommand\'{e}es pour ces
requ\^{e}tes.

\begin{figure}[hbt]
\centering{ \scriptsize{
\begin{tabular}{clcl}\hline

$q_{1}$ & \textbf{select} sales.time\_id, \textbf{sum}(amount\_sold) & $q_{5}$ & \textbf{select} promotions.promo\_name, \\
        & \textbf{from} sales, times & & \textbf{sum}(amount\_sold) \\
        & \textbf{where} sales.time\_id = times.time\_id & &  \textbf{from} sales, promotions\\
        & \textbf{and} times.time\_fiscal\_year = 2000 & &  \textbf{where} sales.promo\_id = promotions.promo\_id\\
        & \textbf{group by}  sales.time\_id  & & \textbf{and} promotions.promo\_begin\_date=`30/01/2000' \\
        & & & \textbf{and} promotions.promo\_end\_date=`30/03/2000' \\
        & & & \textbf{group by} promotions.promo\_name \\ \\

$q_{2}$ & \textbf{select} sales.prod\_id,  & $q_{6}$ & \textbf{select} customers.cust\_marital\_status, \\
        & \textbf{sum}(amount\_sold) & &  \textbf{sum}(quantity\_sold)\\
        & \textbf{from} sales, products, promotions  & & \textbf{from} sales, customers, products\\
        & \textbf{where} sales.prod\_id = products.prod\_id & & \textbf{where} sales.cust\_id = customers.cust\_id \\
        & \textbf{and} sales.promo\_id = promotions.promo\_id & & \textbf{and} sales.prod\_id = products.prod\_id\\
        & \textbf{and} promotions.promo\_category = `news paper' & & \textbf{and} customers.cust\_gender = `woman' \\
        & \textbf{group by} sales.prod\_id & & \textbf{and} products.prod\_name = `shampooing'\\
        & & &  \textbf{group by} customers.cust\_first\_name \\ \\

$q_{3}$ & \textbf{select} customers.cust\_gender, \textbf{sum}(amount\_sold) & $q_{7}$ &  \textbf{select} products.prod\_name, \textbf{sum}(amount\_sold)\\
        & \textbf{from} sales, customers, products, & & \textbf{from} sales, products, promotions \\
        & \textbf{where} sales.cust\_id = customers.cust\_id & & \textbf{where} sales.prod\_id = products. prod\_id \\
        & \textbf{and} sales.prod\_id = products.prod\_id & & \textbf{and} sales.promo\_id =promotions.promo\_id \\
        & \textbf{and} customers.cust\_marital\_status =`single'& & \textbf{and} products.prod\_category=`tee shirt' \\
        & \textbf{and} products.prod\_category = `women' & & \textbf{and} promotions.promo\_end\_date=`30/04/2000' \\
        & \textbf{group by} customers.cust\_gender & & \textbf{group by} products.prod\_name
        \\\\

$q_{4}$ & \textbf{select} products.prod\_name, sum(amount\_sold) & $q_{8}$ & \textbf{select} channels.channel\_desc, sum(quantity\_sold)\\
        & \textbf{from} sales, products, promotions & & \textbf{from} sales, channels\\
        & \textbf{where} sales.prod\_id = products.prod\_id & & \textbf{where} sales.channel\_id = channels.channel\_id \\
        & \textbf{and} sales.promo\_id = promotions.prom\_id & & \textbf{and} channels.channel\_class = `Internet' \\
        & \textbf{and} promotions.promo\_category = `TV' & & \textbf{group by} channels.channel\_desc\\
        & \textbf{group by} products.prod\_name & & \\

\hline
\end{tabular}
}} \caption{Exemple de charge}\label{fig:charge_sim}
\end{figure}

\begin{landscape}

\begin{figure}[hbt]
\centering{ \scriptsize{
\begin{tabular}{clcl}\hline

$v_{1}$ & \textbf{create} \textbf{materialized view} $v_{1}$ \textbf{as} & $v_{5}$ &\textbf{create} \textbf{materialized view} $v_{5}$ \textbf{as}\\
        & \textbf{select} sales.time\_id, times.time\_fiscal\_year, &  &   \textbf{select} sales.prod\_id, products.prod\_category,\\
        & \textbf{sum}(amount\_sold) & &  promotions.promo\_category, \textbf{sum}(amount\_sold) \\
        & \textbf{from} sales, times & &  \textbf{from} sales, products, promotions\\
        & \textbf{where} sales.time\_id = times.time\_id & &  \textbf{where} sales.prod\_id = = products.prod\_id\\
        & \textbf{group by}  sales.time\_id, times.times\_fiscal\_year  & & \textbf{and} sales.promo\_id = promotions.promo\_id\\
        & & & \textbf{group by} sales.prod\_id, products.prod\_category, \\
        & & & promotions.promo\_category \\ \\

$v_{2}$ & \textbf{create} \textbf{materialized view} $v_{2}$ \textbf{as}& $v_{6}$ & \textbf{create} \textbf{materialized view} $v_{6}$ \textbf{as}\\
        & \textbf{select} sales.prod\_id, sales.cust\_id, channels.channel\_desc, & &  \textbf{select} channels.channel\_class, products.prod\_name, channels.channel\_desc,\\
        & channels.channel\_class, \textbf{sum}(quantity\_sold) & & products.prod\_category, \textbf{sum}(sales.quantity\_sold), \textbf{sum}(sales.amount\_sold)\\
        & \textbf{from} sales, channels, products, customers  & & \textbf{from} sales, channels, products \\
        & \textbf{where} sales.prod\_id = products.prod\_id & & \textbf{where} sales.prod\_id = products.prod\_id \\
        & \textbf{and} sales.channel\_id = channels.channel\_id & & \textbf{and} sales.channel\_id = channels.channel\_id\\
        & \textbf{and} sales.cust\_id = customers.cust\_id & &  \textbf{group by} channels.channel\_class, products.prod\_name,\\
        & \textbf{group by} sales.prod\_id, sales.cust\_id, channels.channel\_desc, & & products.prod\_category, channels.channel\_desc\\
        & channels.channel\_class & & \\ \\
$v_{3}$ &\textbf{create} \textbf{materialized view} $v_{3}$ \textbf{as}& $v_{7}$ &  \textbf{create} \textbf{materialized view} $v_{7}$ \textbf{as}\\
        & \textbf{select} customers.cust\_first\_name, products.prod\_name, & & \textbf{select} sales.prod\_id, products.prod\_category, channels.channel\_desc,\\
        & products.prod\_category, customers.cust\_gender, & & promotions.promo\_name, promotions.promo\_begin\_date,  \\
        & customers.cust\_marital\_status, \textbf{sum}(sales.quantity\_sold) & &  promotions.promo\_end\_date, products.prod\_name, \textbf{sum}(sales.quantity\_sold), \\
        & \textbf{from} sales, customers, products & & \textbf{sum}(sales.amount\_sold) \\
        & \textbf{where} sales.cust\_id = customers.cust\_id & &  \textbf{from} sales, products, promotions\\
        & \textbf{and} sales.prod\_id = products.prod\_id & &  \textbf{where} sales.prod\_id = products.prod\_id \\
        & \textbf{group by} customers.cust\_first\_name, products.prod\_name, & &  \textbf{and} sales.promo\_id = promotions.promo\_id \\
        & products.prod\_category, customers.cust\_gender, & & \textbf{and} sales.channel\_id = channels.channel\_id\\
        & customers.cust\_marital\_status & &  \textbf{group by} sales.prod\_id, products.prod\_category, channels.channel\_desc, \\
        & & & promotions.promo\_name, promotions.promo\_begin\_date, \\
        & & & promotions.promo\_end\_date, products.prod\_name\\ \\

$v_{4}$ &\textbf{create} \textbf{materialized view} $v_{4}$ \textbf{as}&  &  \\
        & \textbf{select} products.prod\_name, products.prod\_category, &  &  \\
        & promotions.promo\_category, \textbf{sum}(amount\_sold) & &  \\
        & \textbf{from} sales, products, promotions & & \\
        & \textbf{where} sales.prod\_id = products.prod\_id  & &  \\
        & \textbf{and} sales.promo\_id = promotions.promo\_id & &  \\
        & \textbf{group by} products.prod\_name, products.prod\_category, & & \\
        & promotions.promo\_category &  &  \\ \\

 \hline
\end{tabular}
}} \caption{Vues mat\'{e}rialis\'{e}es candidates}\label{fig:vue_sim}
\end{figure}

\end{landscape}


\begin{figure}[hbt]
\centering{\small{
\begin{tabular}{cl}\hline
\textbf{index} & \textbf{attributs index\'{e}s}\\ \hline
$i_{1}$ & promotions.promo\_category \\
$i_{2}$ & channels.channel\_desc \\
$i_{3}$ & channels.channel\_class \\
$i_{4}$ & customers.cust\_marital\_status \\
$i_{5}$ & customers.cust\_gender \\
$i_{6}$ & times.time\_begin\_date \\
$i_{7}$ & times.time\_end\_date \\
$i_{8}$ & times.fiscal\_year \\
$i_{9}$ & products.prod\_name \\
$i_{10}$ & products.prod\_category \\
$i_{11}$ & promotions.promo\_name \\
$i_{12}$ & customers.cust\_first\_name \\
\hline
\end{tabular}
}}\caption{Index candidats}\label{fig:index_sim}
\end{figure}

\begin{table}[hbt]
\centering
\begin{tabular}{|c|c|c|c|c|c|c|c|}
\cline{2-8}

\multicolumn{1}{c|}{}  & $v_{1}$ & $v_{2}$ & $v_{3}$ &  $v_{4}$ & $v_{5}$ &
$v_{6}$ & $v_{7}$ \\ \hline

$q_{1}$ & 1 & 0 & 0 & 0 & 0 & 0 & 0  \\ \hline

$q_{2}$ & 0 & 0 & 0 & 1 & 0 & 0 & 0  \\ \hline

$q_{3}$ & 0 & 0 & 1 & 0 & 0 & 0 & 0  \\ \hline

$q_{4}$ & 0 & 0 & 0 & 1 & 0 & 0 & 0  \\ \hline

$q_{5}$ & 0 & 0 & 0 & 0 & 0 & 0 & 1  \\ \hline

$q_{6}$ & 0 & 0 & 1 & 0 & 0 & 0 & 0  \\ \hline

$q_{7}$ & 0 & 0 & 0 & 0 & 0 & 0 & 1  \\ \hline

$q_{8}$ & 0 & 1 & 0 & 0 & 0 & 1 & 0  \\ \hline

\end{tabular}
\caption{Matrice requ\^{e}tes-vues}\label{tab:req_vue}
\end{table}

\subsection{Matrice requ\^{e}tes-index}

La matrice requ\^{e}tes-index permet  d'identifier les index
construits sur les tables de la base.  La matrice requ\^{e}tes-index
peut \^{e}tre vue comme la r\'{e}\'{e}criture des requ\^{e}tes de la
charge en fonction des index recommand\'{e}s par un algorithme de
s\'{e}lection d'index. Les lignes et les colonnes de cette matrice
sont respectivement les requ\^{e}tes de la charge et les index
s\'{e}lectionn\'{e}s \`{a} partir de cette charge. Le terme
g\'{e}n\'{e}ral de la matrice est \'{e}gal \`{a} un si une
requ\^{e}te donn\'{e}e exploite un index et \`{a} z\'{e}ro sinon.
Le Tableau~\ref{tab:req_index} illustre un exemple de matrice requ\^{e}tes-index
compos\'{e}e de huit requ\^{e}tes et de douze index recommand\'{e}s pour ces requ\^{e}tes.

\begin{table}[hbt]
\centering
\begin{tabular}{|c|c|c|c|c|c|c|c|c|c|c|c|c|}
\cline{2-13}

\multicolumn{1}{c|}{}  & $i_{1}$ & $i_{2}$ & $i_{3}$ &  $i_{4}$ & $i_{5}$ &
$i_{6}$ & $i_{7}$ & $i_{8}$ & $i_{9}$ & $i_{10}$ &
      $i_{11}$  & $i_{12}$ \\ \hline

$q_{1}$ & 0 & 0 & 0 & 0 & 0 & 0 & 0 & 1 & 0 & 0 & 0 & 0 \\ \hline

$q_{2}$ & 1 & 0 & 0 & 0 & 0 & 0 & 0 & 0 & 0 & 0 & 0 & 0\\ \hline

$q_{3}$ & 0 & 0 & 0 & 1 & 1 & 0 & 0 & 0 & 0 & 1 & 0 & 0 \\ \hline

$q_{4}$ & 1 & 0 & 0 & 0 & 0 & 0 & 0 & 0 & 1 & 0 & 0 & 0 \\ \hline

$q_{5}$ & 0 & 0 & 0 & 0 & 1 & 1 & 0 & 0 & 0 & 0 & 1 & 0 \\ \hline

$q_{6}$ & 0 & 0 & 0 & 0 & 1 & 0 & 0 & 0 & 1 & 0 & 0 & 1 \\ \hline

$q_{7}$ & 0 & 0 & 0 & 0 & 0 & 0 & 1 & 0 & 1 & 1 & 0 & 0 \\ \hline

$q_{8}$ & 0 & 1 & 1 & 0 & 0 & 0 & 0 & 0 & 0 & 0 & 0 & 0 \\ \hline

\end{tabular}
\caption{Matrice requ\^{e}tes-index}\label{tab:req_index}
\end{table}

\subsection{Matrice vues-index}

La matrice vues-index identifie les index construits sur les vues mat\'{e}rialis\'{e}es
recommand\'{e}es par l'algorithme de s\'{e}lection de vues. Les lignes et les colonnes
de cette matrice sont respectivement les vues mat\'{e}rialis\'{e}es candidates
pr\'{e}alablement s\'{e}lectionn\'{e}es et les index candidats recommand\'{e}s pour ces vues.
Le terme g\'{e}n\'{e}ral de la matrice est \'{e}gal \`{a} un si une vue donn\'{e}e exploite un
index et \`{a} z\'{e}ro sinon.
Le Tableau~\ref{tab:vue_index} illustre un exemple de matrice vues-index
compos\'{e}e de sept vues et de douze index recommand\'{e}s pour ces vues.

\begin{table}[hbt]
\centering
\begin{tabular}{|c|c|c|c|c|c|c|c|c|c|c|c|c|}
\cline{2-13}

\multicolumn{1}{c|}{}  & $i_{1}$ & $i_{2}$ & $i_{3}$ &  $i_{4}$ & $i_{5}$ &
$i_{6}$ & $i_{7}$ & $i_{8}$ &  $i_{9}$ & $i_{10}$ & $i_{11}$ & $i_{12}$\\
\hline

$v_{1}$ & 0 & 0 & 0 & 0 & 0 & 0 & 0 & 1 & 0 & 0 & 0 & 0 \\ \hline

$v_{2}$ & 0 & 1 & 0 & 0 & 0 & 0 & 0 & 0 & 0 & 0 & 0 & 0 \\ \hline

$v_{3}$ & 0 & 0 & 0 & 1 & 1 & 0 & 0 & 0 & 1 & 1 & 0 & 1 \\ \hline

$v_{4}$ & 1 & 0 & 0 & 0 & 0 & 0 & 0 & 0 & 1 & 1 & 0 & 0 \\ \hline

$v_{5}$ & 1 & 0 & 0 & 0 & 0 & 0 & 0 & 0 & 0 & 1 & 0 & 0 \\ \hline

$v_{6}$ & 0 & 1 & 1 & 0 & 0 & 0 & 0 & 0 & 1 & 1 & 0 & 0  \\ \hline

$v_{7}$ & 0 & 1 & 0 & 0 & 0 & 1 & 1 & 0 & 1 & 1 & 1 & 0 \\ \hline

\end{tabular}
\caption{Matrice vues-index}\label{tab:vue_index}
\end{table}

\section{Mod\`{e}les de co\^{u}t}\label{sec:cost_modele_sim}

G\'{e}n\'{e}ralement, le nombre d'index et de vues candidats est
d'autant plus important que la charge en entr\'{e}e est volumineuse.
La cr\'{e}ation de tous ces index et vues peut ne pas \^{e}tre
r\'{e}alisable en pratique \`{a} cause de la contrainte d\'{e}finie
sur l'espace de stockage allou\'{e} aux index et vues. Pour pallier
ces limitations, nous exploitons des mod\`{e}les de co\^{u}t
permettant de ne conserver que les index et les vues les plus
avantageux. Ces mod\`{e}les estiment l'espace en octets occup\'{e}
par les index et les vues, les co\^{u}ts d'acc\`{e}s aux donn\'{e}es
\`{a} travers ces index et/ou ces vues et le co\^{u}t de leur
maintenance en terme de nombre d'entr\'{e}es/sorties.

Nous avons d\'{e}velopp\'{e} des mod\`{e}les qui estiment le
co\^{u}t d'acc\`{e}s aux donn\'{e}es \`{a} travers des index
\textit{bitmap} de jointure, ainsi que les co\^{u}ts de maintenance
et de stockage de ces index~\cite{aou05aut}. Nous avons
\'{e}galement pr\'{e}sent\'{e} des mod\`{e}les qui estiment le
co\^{u}t d'acc\`{e}s aux donn\'{e}es \`{a} travers des vues
mat\'{e}rialis\'{e}es, ainsi que les co\^{u}ts de maintenance et de
stockage de ces vues. Dans la suite de la section, nous ne
d\'{e}veloppons donc que les nouveaux mod\`{e}les de co\^{u}t
d\'{e}velopp\'{e}s pour ce travail relatifs aux index en B-arbre.

\section{Calcul du b\'{e}n\'{e}fice de mat\'{e}rialisation et
d'indexation}\label{sec:calcul_benefice}

Le b\'{e}n\'{e}fice apport\'{e} par la s\'{e}lection d'un objet (index, vue mat\'{e}rialis\'{e}e ou vue
mat\'{e}rialis\'{e}e avec index) est d\'{e}fini comme la diff\'{e}rence entre le co\^{u}t des
requ\^{e}tes de la charge \`{a} un moment donn\'{e} et le co\^{u}t de ces m\^{e}mes requ\^{e}tes suite
\`{a} l'ajout de cet objet.

Soient $Q$ une charge de requ\^{e}tes et $Config$ une configuration
compos\'{e}e de vues mat\'{e}rialis\'{e}es et d'index construits sur
les tables de base ou les vues. $QI$, $QV$ et $VI$ sont
respectivement les matrices requ\^{e}tes-index, requ\^{e}tes-vues et
vues-index. Nous d\'{e}veloppons dans les sections suivantes le calcul du
b\'{e}n\'{e}fice pour un index ou une vue mat\'{e}rialis\'{e}e.

\subsection{B\'{e}n\'{e}fice apport\'{e} par un index}

L'ajout d'un index \`{a} la configuration $Config$ peut conduire \`{a} plusieurs
alternatives r\'{e}sum\'{e}es dans le Tableau~\ref{tab:benefice_index}. En effet,
l'ajout d'un index donn\'{e} \`{a} la configuration $Config$ peut am\'{e}liorer de fa\c{c}on
directe le co\^{u}t des requ\^{e}tes de la charge ou indirectement \`{a} travers des vues
auxquelles cet index est associ\'{e}.

\begin{table}[hbt]
\centering{
\begin{tabular}{|c|p{6cm}|c|}
  \cline{2-3}
  \multicolumn{1}{c|}{} & \multicolumn{1}{c|}{$VI[v,i]=1$} & $VI[v,i]=0$ \\ \hline
  $v \in Config$ & min (b\'{e}n\'{e}fice de mat\'{e}rialisation, b\'{e}n\'{e}fice d'indexation de $v$) & b\'{e}n\'{e}fice d'indexation \\ \hline
  $v \notin Config$ & \multicolumn{1}{c|}{\textbf{---}} & b\'{e}n\'{e}fice d'indexation \\ \hline
\end{tabular}
}\caption{B\'{e}n\'{e}fice apport\'{e} par l'ajout d'index}\label{tab:benefice_index}
\end{table}

Le b\'{e}n\'{e}fice d'indexation apport\'{e} par l'ajout d'un index $i$ est calcul\'{e} comme
suit :

\begin{scriptsize}
$$benefice(Q,Config \cup \set{i})= \left\lbrace
\begin{tabular}{ll}
$\frac{C(Q,Config) - C(Q,Config \cup \set{i})}{taille(i)}$ & si $\forall v \in V, VI[v,i]=0$, \\
$\frac{C(Q,Config) - C(Q,Config \cup \set{i} \cup V')}{taille(\set{i})+ \sum_{v'\in V'} taille(\set{v'})}$ & si $V'= \set{v \in Config, VI[v,i]=1} \neq \emptyset$ \\
 0 & sinon.\\
\end{tabular}
\right.$$
\end{scriptsize}

\subsection{B\'{e}n\'{e}fice apport\'{e} par une vue mat\'{e}rialis\'{e}e}

L'ajout d'une vue mat\'{e}rialis\'{e}e \`{a} la configuration
$Config$ peut conduire aux alternatives r\'{e}sum\'{e}es dans le
Tableau~\ref{tab:benefice_vue}. En effet, l'ajout d'une vue
donn\'{e}e \`{a} la configuration $Config$ peut am\'{e}liorer de
fa\c{c}on directe le co\^{u}t des requ\^{e}tes de la charge ou de
fa\c{c}on collaborative avec les index associ\'{e}s \`{a} cette vue.

\begin{table}[hbt]
\centering{
\begin{tabular}{|c|p{6cm}|c|}
  \cline{2-3}
  \multicolumn{1}{c|}{} & \multicolumn{1}{c|}{$VI[v,i]=1$} & $VI[v,i]=0$ \\ \hline
  $i \in Config$ & \multicolumn{1}{c|}{\textbf{---}} & b\'{e}n\'{e}fice d'indexation \\ \hline
  $i \notin Config$ & min (b\'{e}n\'{e}fice d'indexation, b\'{e}n\'{e}fice de mat\'{e}rialisation) & b\'{e}n\'{e}fice de mat\'{e}rialisation \\ \hline
\end{tabular}
}\caption{B\'{e}n\'{e}fice apport\'{e} par l'ajout de vue}\label{tab:benefice_vue}
\end{table}

Le b\'{e}n\'{e}fice de mat\'{e}rialisation apport\'{e} par la vue mat\'{e}rialis\'{e}e $v$ est calcul\'{e} comme suit :

\begin{scriptsize}
$$benefice(Q,Config \cup \set{v})= \left\lbrace
\begin{tabular}{ll}
$\frac{C(Q,Config) - C(Q,Config \cup \set{v})}{taille(v)}$ & si $\forall i \in I, VI[v,i]=0$, \\
$\frac{C(Q,Config) - C(Q,Config \cup \set{v} \cup I')}{taille(\set{v}) + \sum_{i' \in I'} taille(\set{i'})}$ & si $I'= \set{i \in Config, VI[v,i]=1} \neq \emptyset$. \\
 0 & sinon.\\
\end{tabular}
\right.$$
\end{scriptsize}

\section{Algorithme de s\'{e}lection simultan\'{e}e d'index et de vues
mat\'{e}rialis\'{e}es}\label{sec:algo_sim}

Notre algorithme de s\'{e}lection d'index et de vues mat\'{e}rialis\'{e}e
(Algorithme~\ref{algo:sel_vue_ind}) est bas\'{e} sur une recherche gloutonne dans
l'ensemble $O$ des objets obtenus en r\'{e}alisant l'union de l'ensemble d'index
candidats $I$ et de l'ensemble de vues candidates $V$ ($O=I \cup V$).
Soir la fonction objectif $F$ d\'{e}finie comme suit :

$$F_{/Config}(\set{o_{i}})=\textrm{b\'{e}n\'{e}fice}(Q,Config\cup (\set{o_{i}}))-\beta
C_{maitenance}(\set{o_{i}})$$ o\`{u} $o_{i}$ peut \^{e}tre un index
ou une vue et $\beta = |Q| \: p(o_{i})$ estime le nombre de mises
\`{a} jour de  $o_{i}$. La probabilit\'{e} de mise \`{a} jour
$p(o_{i})$ est \'{e}gale \`{a} $\frac{1}{\textrm{nombre
d'\'{e}l\'{e}ments de
}O}\frac{\%\textrm{rafra\^{\i}chissement}}{\%\textrm{interrogation}}$,
o\`{u} le ratio
$\frac{\%\textrm{rafra\^{\i}chissement}}{\%\textrm{interrogation}}$
repr\'{e}sente la proportion de rafra\^{\i}chissement par rapport
\`{a} la proportion d'interrogation de l'entrep\^{o}t de
donn\'{e}es.

\begin{algorithm}[!h]
\caption{S\'{e}lection simultan\'{e}e de vues mat\'{e}rialis\'{e}es et d'index}
\label{algo:sel_vue_ind}
\begin{algorithmic}[1]

\STATE $Config \leftarrow \emptyset$

\STATE $O \leftarrow I \cup V$

\REPEAT

    \STATE $o_{max} \leftarrow \emptyset$
    \STATE $taille(o_{max}) \leftarrow 0$
    \STATE $benefice_{max} \leftarrow 0$

    \FORALL{$o_{i} \in O-Config$}
        \IF{$F_{Config}(\set{o_{i}})>benefice_{max}$}
            \STATE $benefice_{max} \leftarrow F_{Config}(\set{o_{i}})$
            \STATE $o_{max} \leftarrow \arg\max(F_{Config}(\set{o_{i}})$
            \STATE \COMMENT{l'ensemble $o_{max}$ peut contenir des index et des vues mat\'{e}rialis\'{e}es}
        \ENDIF
    \ENDFOR
    \IF{$F_{Config}(o_{max}) > 0$}
        \IF{$index(o_{max})=vrai$}
            \STATE $taille(o_{max}) \leftarrow taille_{index}(o_{max})$
        \ELSE
            \IF{$vue(o_{max})=vrai$}
                \STATE $taille(o_{max}) \leftarrow taille_{vue}(o_{max})$
            \ELSE
                \STATE $o_{max} = i_{max} \cup v_{max}$
                    \STATE $taille(o_{max}) \leftarrow taille(o_{max}) + taille_{index}(i_{max})+ taille_{vue}(v_{max})$
            \ENDIF
        \ENDIF
    \ENDIF
    \STATE $S \leftarrow S - taille(o_{max})$
    \STATE $Config \leftarrow Config \cup (o_{max})$
\UNTIL{($F_{Config}(o_{max})\leq 0$ ou $O-Config = \emptyset$ ou $S
\leq0$)}
\end{algorithmic}
\end{algorithm}

Soit $S$ l'espace disque allou\'{e} par l'administrateur de
l'entrep\^{o}t de donn\'{e}es pour stocker les vues
mat\'{e}rialis\'{e}es et les index. \`{A} la premi\`{e}re
it\'{e}ration de notre algorithme, les valeurs de la fonction
objectif $F$ sont calcul\'{e}es pour chaque index ou vue de
l'ensemble $O$. Le co\^{u}t d'ex\'{e}cution de toutes les
requ\^{e}tes de la charge $Q$ est \'{e}gal au co\^{u}t
d'ex\'{e}cution de ces requ\^{e}tes \`{a} partir des tables de base
sans indexation ni vue. L'ensemble $o_{max}$ de vues et/ou d'index
qui maximise $F$, s'il existe ($F_{Config}(\set{o_{max})}>0$), est
alors ajout\'{e} \`{a} l'ensemble $Config$ si l'espace de stockage
$S$ n'est pas atteint. Si l'ajout est fructueux, l'espace $S$ est
diminu\'{e} de l'espace de stockage occup\'{e} par $o_{max}$.
L'espace occup\'{e} par $o_{max}$ d\'{e}pend de son contenu (index
et/ou vue). Nous utilisons les fonctions bool\'{e}ennes
$index(o_{max})$ et $vue(o_{max})$ qui renvoient vrai si $o_{max}$
est un index ou une vue (lignes~15~\`{a}~25 de
l'Algorithme~\ref{algo:sel_vue_ind}), respectivement.

Les valeurs de la fonction objectif $F$ sont ensuite recalcul\'{e}es pour chaque
\'{e}l\'{e}ment restant dans $O-Config$, car elles d\'{e}pendent des vues et des index
s\'{e}lectionn\'{e}s pr\'{e}sents dans $Config$. C'est cela qui permet de prendre en compte
les interactions qui peuvent exister entre les index et les vues mat\'{e}rialis\'{e}es.
Rappelons que cette interaction est implicitement inclue dans le calcul du
b\'{e}n\'{e}fice, qui exploite les matrices requ\^{e}tes-index $QI$, requ\^{e}tes-vues $QV$ et
vues-index $VI$.

Nous r\'{e}p\'{e}tons ces it\'{e}rations jusqu'\`{a} ce qu'il n'y
ait plus d'am\'{e}lioration de la fonction objectif
($F_{/Config}(o_{max}) \leq 0$), que tous les index et vues aient
\'{e}t\'{e} s\'{e}lectionn\'{e}s ($O-Config= \emptyset $) ou que la
limite d'espace de stockage soit atteinte ($S \leq 0$).

\section{Exp\'{e}rimentations}\label{sec:exp_sim}

Afin de valider notre strat\'{e}gie de s\'{e}lection simultan\'{e}e d'index et de vues
mat\'{e}\-rialis\'{e}es, nous l'avons exp\'{e}riment\'{e}e sur un entrep\^{o}t de donn\'{e}es test
implant\'{e} au sein du SGBD Oracle~9i. Nos exp\'{e}rimentations ont \'{e}t\'{e} r\'{e}alis\'{e}es sur
un PC sous Windows~XP~Pro dot\'{e} d'un processeur Pentium 4 \`{a} 2.4~GHz, d'une
m\'{e}moire centrale de 512~Mo et d'un disque dur IDE de 120~Go.

Notre entrep\^{o}t de donn\'{e}es test est compos\'{e} d'une table de faits
\textbf{\texttt{Sales}} et de cinq tables dimensions
\texttt{\textbf{Customers}}, \textbf{\texttt{Products}},
\textbf{\texttt{Promotions}}, \textbf{\texttt{Times}} et
\textbf{\texttt{Channels}}. Le Tableau~\ref{tab:nb_taile}
d\'{e}taille le nombre de n-uplets et la taille en Mo de chacune des
tables de cet entrep\^{o}t.
\begin{table}[hbt]
\begin{center}
\begin{tabular}{|l|r|r|} \hline
\textbf{Table} & \textbf{Nombre de n-uplets} & \textbf{Taille (Mo}) \\
\hline \hline Sales & \nombre{16 260 336} & \nombre{372,17}
\\ \hline Customers & \nombre{50 000} & \nombre{6,67} \\ \hline
Products & \nombre{10 000} & \nombre{2,28} \\ \hline Times & \nombre{1 461} & \nombre{0,20} \\
\hline Promotions & \nombre{501} & \nombre{0,04} \\ \hline Channels & \nombre{5} & \nombre{0,0001} 
\\
\hline
\end{tabular}
\caption{Caract\'{e}ristiques de l'entrep\^{o}t de donn\'{e}es test}\label{tab:nb_taile}
\end{center}
\end{table}

\begin{figure}[hbt]
\begin{center}
{\centering \resizebox*{0.9\textwidth}{!}{\includegraphics{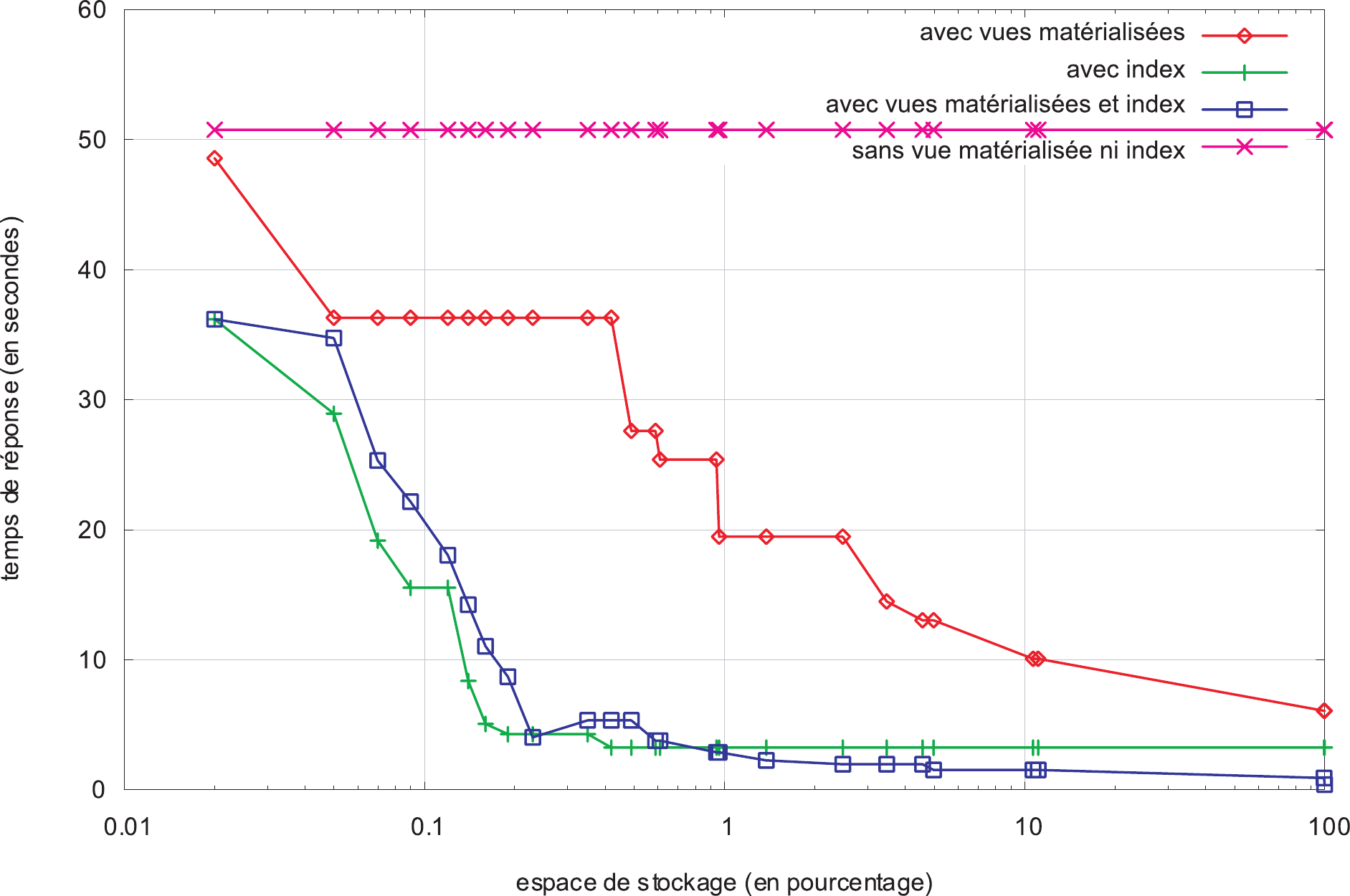}}
\par}
\caption{R\'{e}sultats exp\'{e}rimentaux}\label{fig:resultat_vues_index}
\end{center}
\end{figure}


Nous avons mesur\'{e} le temps d'ex\'{e}cution des requ\^{e}tes de
la charge de la Figure~\ref{fig:charge_sim} dans les cas suivants : sans index ni
vue mat\'{e}rialis\'{e}e, avec vues mat\'{e}rialis\'{e}es, et avec
index et vues mat\'{e}rialis\'{e}es. La
Figure~\ref{fig:resultat_vues_index} repr\'{e}sente la variation de
ce temps de r\'{e}ponse en fonction du pourcentage d'espace de stockage
utilis\'{e}. Ce pourcentage est calcul\'{e} par rapport \`{a}
l'espace total occup\'{e} par tous les index et les vues
obtenus en appliquant notre strat\'{e}gie sans d\'{e}finir aucune contrainte d'espace.
Afin de  mieux visualiser les r\'{e}sultats, nous
avons utilis\'{e} une \'{e}chelle logarithmique sur l'axe des
abscisses.

La Figure~\ref{fig:resultat_vues_index} montre que pour les valeurs
\'{e}lev\'{e}es de l'espace de stockage, la s\'{e}lection
simultan\'{e}e d'index et de vues est meilleure que la s\'{e}lection
isol\'{e}e des index et vues. En revanche, nous constatons que pour
les petites valeurs de l'espace de stockage, il peut arriver que la
s\'{e}lection d'index soit plus performante que la s\'{e}lection
simultan\'{e}e d'index et de vues. Cela  peut \^{e}tre expliqu\'{e}
par le fait qu'en g\'{e}n\'{e}ral, la taille des index est
significativement moins importante que celle des vues. Dans ce cas,
on peut mettre dans le m\^{e}me espace (de petite taille) plus
d'index que de vues et ainsi am\'{e}liorer davantage le temps de
r\'{e}ponse. En effet, intuitivement, plus on dispose d'index, plus on am\'{e}liore les
performances. Or, ce point n'avait pas \'{e}t\'{e} mis en lumi\`{e}re par les travaux ant\'{e}rieurs aux n\^{o}tres.

\section{Conclusion et perspectives}\label{sec:conclusion_sim}

Nous avons propos\'{e} dans cet article une nouvelle d\'{e}marche de s\'{e}lection
simultan\'{e}e d'index et de vues mat\'{e}rialis\'{e}es dans les entrep\^{o}ts de donn\'{e}es.
Notre approche prend r\'{e}ellement en compte l'interaction qui peut exister entre
les index et les vues mat\'{e}rialis\'{e}es et les traite simultan\'{e}ment afin de r\'{e}duire
le co\^{u}t d'ex\'{e}cution de requ\^{e}tes. Cela donne lieu \`{a} une s\'{e}lection optimale des
vues et des index en fonction de l'espace de stockage qui leur est allou\'{e}. En
effet, nos r\'{e}sultats exp\'{e}rimentaux montrent que la s\'{e}lection simultan\'{e}e de vues
mat\'{e}rialis\'{e}es et d'index est plus performante que la s\'{e}lection isol\'{e}e de ces
structures lorsque l'espace de stockage est raisonnablement grand.

Les perspectives ouvertes par ces travaux sont de plusieurs ordres. Tout d'abord, il est
n\'{e}cessaire de poursuivre nos exp\'{e}rimentations afin de confronter
notre approche \`{a} l'existant. Malheureusement, la proposition
\textit{joint enumeration} d'Agrawal~\etal, qui est
la plus proche de la n\^{o}tre, n'est pas suffisamment document\'{e}e pour que
nous puissions mener une telle \'{e}tude \`{a} bien. En revanche, comparer nos travaux \`{a} ceux
de Bellatreche~\etal, \`{a} la fois en termes de gains de performance et de surcharge pour
le syst\`{e}me, pourrait nous permettre d'\'{e}tablir d\'{e}finitivement que
la s\'{e}lection conjointe d'index et de vues mat\'{e}rialis\'{e}es est plus
int\'{e}ressante que leur s\'{e}lection concurrente.

Par ailleurs, cette strat\'{e}gie d'optimisation des performances s'applique dans un cas statique.
La charge sur laquelle est effectu\'{e}e l'optimisation peut devenir
obsol\`{e}te au bout d'un temps donn\'{e}. Lorsque cela arrive, il
faut res\'{e}lectionner les index et les vues
mat\'{e}rialis\'{e}es. Les travaux traitant de la d\'{e}tection de sessions bas\'{e}s
sur le calcul d'entropie~\cite{qui05mac} pourraient s'appliquer pour
d\'{e}terminer le moment o\`{u} il faut lancer la res\'{e}lection
des index et des vues.

Dans ces travaux, nous nous sommes \'{e}galement
positionn\'{e}s dans le cas o\`{u} l'administrateur optimise les
requ\^{e}tes adress\'{e}es au syst\`{e}me par tous les utilisateurs confondus. Or, les
besoins d\'{e}finis par diff\'{e}rents profils d'utilisateurs (dans le cas des syst\`{e}mes
multi-utilisateurs) sont diff\'{e}rents. Il serait donc plus pertinent d'appliquer
nos strat\'{e}gies sur des groupes de requ\^{e}tes d\'{e}finis par les utilisateurs
identifi\'{e}s dans chaque profil.

Finalement, nous nous sommes restreints \`{a} utiliser seulement les index et les
vues mat\'{e}rialis\'{e}es comme m\'{e}canismes d'optimisation des performances. Or,
d'autres structures de donn\'{e}es peuvent \^{e}tre int\'{e}gr\'{e}es ou coupl\'{e}es avec les
index et les vues, comme par exemple la gestion de cache, le regroupement et le
partitionnement (fragmentation)~\cite{agrawal04,zilio04,bel05evo}.

\bibliographystyle{RNTIBiblio}
\bibliography{eda06}

\end{document}